\documentclass[runningheads]{llncs}
\usepackage[english]{babel}
\usepackage[nottoc]{tocbibind}
\usepackage{parskip}
\title{Bibliography management: BibTeX}
\author{Overleaf}
\usepackage{setspace}
\usepackage{graphicx}
\begin{document}

\title{Lung-CADex: Fully automatic Zero-Shot Detection and Classification of Lung Nodules in Thoracic CT Images\thanks{Supported by Higher Education Commission Pakistan under NRPU Project no:17019}}
\titlerunning{Lung-CADex}
%
%
\author{Furqan Shaukat$^{1,2}$, Syed Muhammad Anwar $^{3,4}$, Abhijeet Parida $^3$, Van Khanh Lam $^3$,
Marius George Linguraru $^{3,4}$, Mubarak Shah $^2$}
\authorrunning{Shaukat et al.}
%
\institute{ $^1$Faculty of Electronics and Electrical Engineering, University of Engineering and Technology, Taxila 47080, Pakistan. $^2$ Center for Research in Computer Vision, University of Central Florida, USA. $^3$ Sheikh Zayed Institute for Pediatric Surgical Innovation, Childrens National Hospital, Washington DC. $^4$ School of Medicine and Health Sciences, George Washington University, Washington DC\\
\email{furqan.shaukat@ucf.edu}
}
\maketitle              

\begin{abstract}
Lung cancer has been one of the major threats to human life for decades. Computer-aided diagnosis can help with early lung nodule detection and facilitate subsequent nodule characterization. Large Visual Language models (VLMs) have been found effective for multiple downstream medical tasks that rely on both imaging and text data. However, lesion level detection and subsequent diagnosis using VLMs have not been explored yet. 
We propose CADe, for segmenting lung nodules in a zero-shot manner using a variant of the Segment Anything Model called MedSAM. CADe trains on a prompt suite on input computed tomography (CT) scans by using the CLIP text encoder through prefix tuning. We also propose, CADx, a method for the nodule characterization as benign/malignant by making a gallery of radiomic features and aligning image-feature pairs through contrastive learning. Training and validation of CADe and CADx have been done using one of the largest publicly available datasets, called LIDC. To check the generalization ability of the model, it is also evaluated on a challenging dataset, LUNG\textsubscript{x}. Our experimental results show that the proposed methods achieve a sensitivity of 0.86 compared to 0.76 that of other fully supervised methods.The source code, datasets and pre-processed data can be accessed using the link:



\keywords{Lung Nodule Detection  \and Computer Aided Diagnosis \and Malignancy prediction.}
\end{abstract}
\vspace*{-\baselineskip}
\section{Introduction}
\vspace*{-\baselineskip}
Lung cancer is one of the most commonly occurring cancers worldwide, with around 2.2 million new cases recorded in 2020~\cite{sung2021global}. Approximately 225,000 people are diagnosed with lung cancer every year in the United States costing \$12 billion~\cite{siegel2019cancer,mariotto2011projections}. A report published by the European Society for Medical Oncology (ESMO) ~\cite{planchard2018metastatic}, indicates that the highest incidence of lung cancer is found in central/eastern European and Asian populations. The situation in developing countries (such as in Asia) is particularly dire ~\cite{moore2010cancer}. Studies have shown that the survival rate can be significantly improved by early detection of lung nodules ~\cite{siegel2019cancer}. However, detecting lung cancer at an early stage is challenging due to 1) a lack of symptoms in most patients 2) an extensive amount of data in terms of computed tomography (CT) scans, and 3) the interobserver variability in nodule detection. Computer-aided diagnosis (CAD) can help with early lung nodule detection and its subsequent nodule characterization. Generally, lung nodules have been characterized as the primary symptom of lung cancer, forming within and in the peripheries of the lungs. In radiology, CAD systems assist clinical experts in the analysis of medical images~\cite{shaukat2019computer}, to detect and localize structures of interest in a semi- or fully-automated manner. 

With the five-year survival rate for patients diagnosed with lung cancer being the lowest among all other cancers, there is a clear need for the design of automated and robust systems that facilitate its early detection, diagnosis, and treatment. A lot of research has been done on developing a lung nodule detection system using conventional machine learning and deep learning techniques ~\cite{tang2019nodulenet,zhu2018deeplung,setio2016pulmonary,liu2020no,niu2022unsupervised}. However, the availability of labeled data has been a major bottleneck for the generalization of these methods. Since, manually annotating this extensive amount of data present in CT scans is laborious and requires trained personnel. In addition, the diagnosis part has not been investigated that much, even though only a robust end-to-end system coupled with detection and diagnosis can be useful in real-time clinical scenarios ~\cite{ozdemir20193d}. 

With the recent advent of large visual language models (VLMs) and their ability to generalize to unseen tasks, there has been an effort within the healthcare community to adapt them to various medical downstream tasks. Specifically, the Segment Anything Model (SAM) ~\cite{kirillov2023segment} has performed exceptionally well on different segmentation tasks on normal real-world images. The domain gap present in these real-world and medical (such as radiology) images and its adaptability have been investigated in different variants of SAM ~\cite{ma2024segment,qiu2023learnable,gong20233dsam,shaharabany2023autosam}. However, these variants have largely been investigated at the anatomical level, and the lesion-level detection and subsequent characterization have not been investigated yet. 

To solve the problem we propose solutions for both computer aided detection (CADe) and computer aided diagnosis (CADx) for lung cancer. These methods effectively segment nodules in a CT scan and can automatically generate the radiomic features associated with the lung nodules to be used to classify if a nodule is benign or malignant. 
\textbf{Our contribution} can be summarized as follows:
\begin{enumerate}
\item Development of a fully automatic end-to-end pipeline for lung cancer diagnosis.
\item Zero-shot detection and segmentation of nodules from lung CT images.
\item Design of a textual prompt suite and adaptation of MedSAM via prefix tuning.
\item Subsequent characterization of segmented lung nodules into benign/malignant via image-feature contrastive learning.
\end{enumerate}
\vspace*{-\baselineskip}

\section{Methodology}
\subsection{Dataset Curation and Pre-Processing} 
We have used the Lung Image Database Consortium (LIDC) dataset ~\cite{armato2011lung} for training and validation purposes. The LIDC data contains 1018 scans, along with nodule annotations by four expert radiologists in a double-blind fashion. We have considered a subset of LIDC called LUNA ~\cite{setio2017validation}, consisting of 888 scans, which removes the inconsistent cases from the original dataset. The nodule inclusion criteria of LUNA have been followed in subsequent nodules' evaluations which gives a total of 1186 nodules. In addition to the nodule annotations, each radiologist has provided the radiological assessment of the respective nodules. Each nodule has been scored on a scale of 1–5 with respect to different radiological features namely subtlety, internal structure, roundness/sphericity, calcification, margin, lobulation, spiculation, and internal texture. The malignancy rating (1, 2, 3, 4, and 5) of the nodules is also given. We have taken four sets of values (by four radiologists) and averaged them to make a single reading for each nodule patch. For training purposes in the diagnosis (CADx) part, we have segmented the nodule patches from their median slice for their ground truth. We have made a gallery of the radiomic features of each nodule patch, which is being used as an input for the diagnosis model along with the nodule patches. 
\vspace*{-\baselineskip}
\subsection{Network Architecture} 
The block diagram of our proposed model is shown in Figure 1. Our proposed method consists of two stages. The first stage consists of the nodule detection module, whereas the second stage consists of the diagnosis module.
\begin{figure}
\centering
\includegraphics[width=0.7\textwidth]{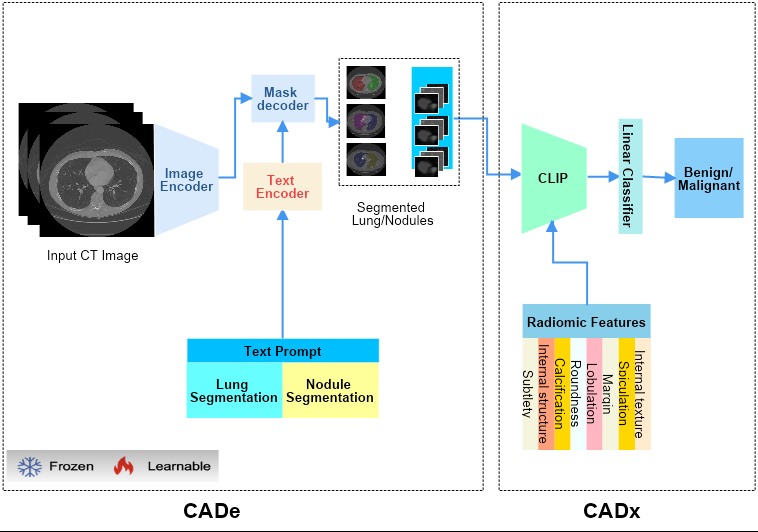}
\caption{Flow chart of the proposed method which consists of two stages namely $CAD_{e}$ which refers to the detection phase and $CAD_{x}$ which refers to the diagnosis phase.} \label{fig1}
\end{figure}
\paragraph{\textbf{Nodule Detection (CADe):}}  For the detection module, we have used MedSAM ~\cite{ma2024segment} in a zero-shot manner via prefix tuning and replaced its visual prompt with a textual prompt. The rationale behind doing this is to enforce the concept of fully automatic analysis for this downstream task of identification and classification of lung nodules. In its current setting, MedSAM provides excellent segmentation results; however, its semi-automatic nature can add the time and effort required for traversing through a complete scan, which can hamper the concept of computer-aided detection in real-time clinical scenarios. For example, if the radiologist or clinician has to provide the bounding box for each slice of CT by first looking into the targeted areas, then its utility for this specific downstream task remains limited. To overcome this challenge, in our proposed strategy we replace the visual prompt with the textual prompt suite. 

With our modified architecture, a clinician or radiologist can see the segmented region of interest (i.e., lung/nodule) by giving textual prompts to the model, respectively. Keeping in view the nature of the task, we have trained the text encoder and fine-tuned it with our textual prompts, while the image encoder and mask decoder have been kept frozen. For quick reference, the image encoder (\texttt{ViT\_Base: 12 layers}) transforms the input image into a high-dimensional image embedding space. The prompt encoder converts the user-provided textual prompt into feature representations using positional encoding. The mask decoder (\texttt{Lightweight: 2 Transformer layers}) combines the image embedding and prompt features through cross-attention. We would like to highlight that we have used a 2D MedSAM model here because of its application in the diagnosis pipeline, where slice-by-slice traversing can be more deterministic in the subsequent characterization of nodules. During the training phase, the model is trained with text-image pairs, and during inference, the model can identify and segment the nodules present in respective CT slices. 
\vspace*{-\baselineskip}
\paragraph{\textbf{Nodule Classification (CADx):}} The second stage of our proposed method consists of the diagnosis module. Leveraging the exceptional alignment power of CLIP ~\cite {radford2021learning}, we have treated the classification task as a retrieval task. We introduce incremental novelties to the CLIP model with modifications to efficiently perform nodule classification. In particular, the overreaching idea is to form a gallery of radiomic features associated with segmented nodule patches and align the model with text (radiomic features)-image (nodule patch) pairs. Once the model is trained, during inference, the model should output the most similar class from the radiomic feature gallery against the given nodule patch. Finally, this is then fed to a linear classifier for binary classification, i.e., benign or malignant. The rationale behind using this model rather than inputting the radiomic features directly into a linear classifier is to leverage the learned representations of the model, which can significantly improve the classification performance. We have used the Resnet50 ~\cite{he2016deep} as an image encoder. To this end, the radiomic features are fed directly to the projection head, and the nodule patch is input to an image encoder. Both of these feature representations are then fused, a cosine similarity matrix is computed, and the most similar class is given as an output, which is finally fed to a linear classifier for final classification. Figure 2 shows the detailed architecture of the diagnosis part.  

Overall, we develop a complete end-to-end pipeline, that includes both detection and diagnosis. One advantage of this strategy is that any false positives during the detection stage would be reduced or eliminated at the diagnosis stage. To ensure this, during the detection phase, we kept the precision high so that the model would not miss any potential nodules, and the false positives were reduced at the diagnosis stage. 
\vspace*{-\baselineskip}
\begin{figure}
\centering
\includegraphics[width=0.7\textwidth]{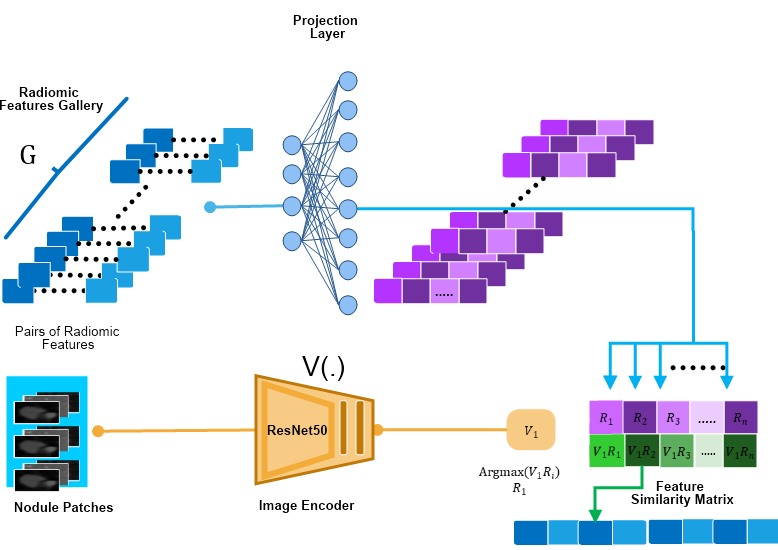}
\caption{Detailed architecture of the lung nodule classification method.} \label{fig2}
\end{figure}
\vspace*{-\baselineskip}
\subsection {Loss Function}

\paragraph{\textbf{Nodule Segmentation:}} 
For the task of nodule segmentation, we have used the same loss function as in MedSAM. For reference, the unweighted sum of cross-entropy loss and dice loss were selected because of their robustness and wide adaptation in medical image segmentation tasks. Specifically, let $S$ and $G$ represent the segmentation result and ground truth, respectively. $s_i$, $g_i$ represent the predicted segmentation and ground truth of voxel $i$, respectively. $N$ is the number of voxels in the image $I$, the binary cross-entropy loss is given as:

\begin{equation}
   L_{BCE} = -\frac{1}{N}\sum_{i=1}^{N}[g_ilogs_i+(1-g_i)log(1-s_i)],
\end{equation}

and dice loss is defined by

\begin{equation}
   L_{Dice} = 1-\frac{2\sum_{i=1}^Ng_is_i}{\sum_{i=1}^N(g_i)^2+\sum_{i=1}^N(s_i)^2},
\end{equation}

The final loss $L$ is defined by

\begin{equation}
   L = L_{BCE} + L_{Dice},  
\end{equation}
\vspace*{-\baselineskip}
\paragraph{\textbf{Nodule Classification:}}
For nodule classification as benign or malignant, we used the loss function defined in CLIP~\cite {radford2021learning} as:

\begin{equation}
L_{SCE} = \alpha \cdot L_{CE}(p, q) + \beta \cdot L_{RCE}(p, q),
\label{lossfunc}
\end{equation}

where:
\begin{itemize}
    \item $L_{CE}(p, q) = -\sum_{i} p_i \log(q_i)$ denotes the cross-entropy loss,
    \item $L_{RCE}(p, q) = -\sum_{i} q_i \log(p_i)$ denotes the reverse cross-entropy loss,
    \item $\alpha$ and $\beta$ are the weighting coefficients for the cross-entropy loss and the reverse cross-entropy loss, respectively,
    \item $p_i$ and $q_i$ represent the predicted probabilities and the true probabilities, respectively.
\end{itemize}
\vspace*{-\baselineskip}
\section{Experimental Results and Discussion}
The experimental setup consists of two parts. In the first phase, the detection model was trained to highlight areas within the input CT image that contain nodules, and hence segment the nodule patches using a text prompt. In the second phase, the segmented nodule patches were fed to the diagnosis model for final prediction as benign/malignant. 

For the nodule detection model, we prefix-tune the text encoder of the MedSAM with \textit{"nodules", "nodule", "lung nodule", "LUNG NODULE", "Nodule", "segment nodule"} as prompts to generate nodule segmentation. 70\% of the LIDC scans were used for prefix-tuning and the rest were reserved for inference testing. The CTs are converted to slices and normalized with window level 40 and window width 400. The MedSAM model was optimized to minimize Dice and BCE loss using AdamW optimizer with a learning rate $5e^{-5}$ and batch size of 16 for 500 epochs. 
\begin{figure}
\centering
\includegraphics[width=0.75\textwidth]{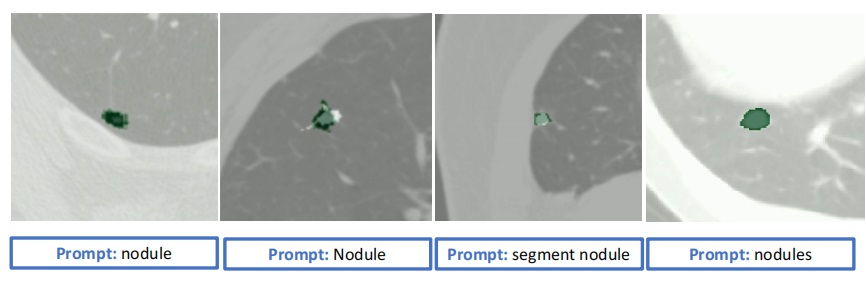}
\caption{Few examples of zero-shot segmentation results generated using the segmentation model and corresponding textual prompts. Green represents the segmented regions, and white represents the ground truth. } \label{medsam}
\end{figure}
We train the diagnosis model using contrastive loss between image-radiomic feature pairs. A feature gallery was created using the average radiomic assessments of four expert radiologists for each nodule patch.
The median nodule slice from the CT was resized to 96x96 to generate an image pair for the radiomic features. We train the  Resnet50 image encoder and the radiomic projection layer to optimize for increased feature similarity. The image embedding dimension of 2048 and a radiomic feature embedding dimension of 8 were used. The model was trained to minimize the loss function from equation \ref{lossfunc} for 500 epochs using a batch size of 8. 

During inference, the output of the detection phase, i.e., the segmented nodule patches, was given as input to the diagnosis model, and the output was then fed to a linear classifier for binary classification i.e., benign/malignant test nodule patches. Malignancy label (scaled 1-5) was used to generate a ground truth for the samples by thresholding. Sample scores greater than three were considered malignant. Samples with an average malignancy value of three were left out during the evaluation. 

We have done our evaluations on two different datasets 1) 30\% holdout samples of LIDC and 2) LUNG\textsubscript{x} \cite{armato2016lungx} which contains hard malignancy labels (pathology proven) of 73 nodules. Standard performance metrics, namely area under the curve, sensitivity, accuracy, and specificity, were used for evaluation. We have compared our test results with two notable recent studies \cite{causey2018highly,choi2022cirdataset}. However, we would like to note that this is not a one-to-one comparison, keeping in mind the semi-automatic and fully supervised nature of the other two studies. The test results of the two studies reported by \cite{choi2022cirdataset} along with ours are shown in Table 1 for comparison. It can be seen that our proposed method performs at par with these fully supervised methods. For the LIDC test dataset, our method outperforms other methods in terms of sensitivity and accuracy, even though our test dataset consists of 264 nodules (4 times large)  as compared to their results, which were reported on a dataset containing 72 nodules with pathology-given labels. For this, we selected LUNG\textsubscript{x} dataset, which contains 60 contrast-enhanced CT scans, with 73 pathology-proven nodules with their hard labels. Keeping in view the notion of zero-shot learning, we do not use the 10 CT scans of this dataset given for calibration which might be a factor in a slight drop of AUC and sensitivity as compared to the LIDC test dataset shown in Table 1. However, the overall results on this dataset show the ability of our method to generalize even to unseen data coming from different image acquisition protocols and with other parameters. Our method outperforms \cite{choi2022cirdataset} in terms of accuracy and specificity and performs at par with \cite{causey2018highly} in terms of AUC for LUNG\textsubscript{X}. 

\noindent{\textbf{Ablation Studies:}} Since we can query top-k radiomic feature matches from CADe, we conduct an ablation study for the 'k' radiomic feature required for a good malignancy prediction. The results have been summarized in Table 2. It can be seen that k=5 (our picked model) gives us the best sensitivity compared to others, which reflects that our diagnosis model performs five-shot learning. In addition, \textbf{k=9} achieves the best \textbf{AUC of 0.81} which outperforms all other baselines, but due to its lower sensitivity, we have not picked that model as the best-performing model. 

\begin{table}[t]
\centering
\caption{Comparison of malignancy prediction metrics, \textbf{N} stands for number of nodules. AUC: area under receiver operating characteristic curve.}
\begin{tabular}{ l|c|c|c|c } 
\hline
\multicolumn{5}{c}{\textbf{Test results on LIDC dataset }} \\
\hline
Network& AUC &Accuracy &Sensitivity &Specificity \\
\hline

CIRD\cite{choi2022cirdataset} (N=72) &0.73 &0.68 &0.81 &0.57 \\
LungX\cite{causey2018highly} (N=72) &0.68 &0.68 &0.78 &0.55 \\
\textbf{Ours(N=264)} &0.69 &\textbf{0.71} & \textbf{0.86} &0.56 \\
\hline

\multicolumn{5}{c}{\textbf{Test results on LUNGx (N = 73)}}\\
\hline
Network &AUC &Accuracy &Sensitivity &Specificity\\
\hline
LungX\cite{causey2018highly} &0.670 &- &- &- \\
CIRD\cite{choi2022cirdataset} &0.733 &68.49 &80.56 &56.76 \\
\textbf{Ours} & 0.656 & \textbf{70.59} & 66.67 & \textbf{73.33} \\
\hline
\end{tabular}
\end{table}

\begin{table}[t]
\centering
\caption{Comparison of number of learning examples used for CLIP contrastive learning. AUC: area under receiver operating characteristic curve, ACC: accuracy.}\label{tab2}
\resizebox{0.8\columnwidth}{!}{%
\begin{tabular}{l|ccccccccc}
\hline
                     & \multicolumn{9}{c}{Number of learning examples $(k)$}                                    \\ \cline{2-10} 
 & 1 & 2 & 3& 4 & 5 & 6 & 7 & 8 & 9 \\ \hline
AUC         & 0.621 & 0.721 & 0.765 & 0.698  & \textbf{0.698} & 0.748 & 0.763 & 0.792 & 0.810 \\
Sensitivity & 86.67 & 73.33 & 66.67 & 100.00 & \textbf{86.77} & 80.00 & 66.67 & 66.67 & 66.67 \\
Specificity & 37.50 & 68.62 & 75.00 & 50.00  & \textbf{56.33} & 56.33 & 62.50 & 62.50 & 62.50 \\
F1         & 68.40 & 70.96 & 68.95 & 78.94  & \textbf{74.25} & 70.58 & 64.50 & 64.50 & 64.50 \\
ACC       & 61.29 & 70.96 & 70.96 & 74.16  & \textbf{70.96} & 67.74 & 64.51 & 64.50 & 64.50 \\ \hline
\end{tabular}%
}
\end{table}

\noindent{\textbf{Model limitations:}} One of the main limitations of this work is the limited annotated data which would have affected the system's overall performance. Another limitation could be the reliance on weak labels for classification phase. With these large foundation models, large, balanced, and well-annotated/strong-labeled data can certainly increase the performance of the system. Another direction for future work can be the integration of Electronic Medical Records into the pipeline for final diagnosis to fully exploit the potential of these multimodal models.
\vspace*{-\baselineskip}
\section{Conclusions}
In this paper, we present an end-to-end pipeline using both CADe and CADx for lung nodule detection and its subsequent malignancy characterization. We used a variant of the Segment Anything Model called MedSAM in a zero-shot manner for the detection part and a CLIP model for further characterization of nodules into benign and malignant. We replaced the visual prompts of MedSAM with textual prompts and designed a prompt suite for this specific downstream task. After segmenting the nodule patches, we formed a radiomic feature gallery and trained the modified CLIP model with nodule patches and their associated radiomic feature sets. During inference, the model gave the most similar class fed to a linear classifier for the final decision. Our results have shown significant value in the detection of large and diverse data. The proposed tool can be used for early lung cancer screening in an end-to-end manner.
\vspace*{-\baselineskip}

%
%
%


 \bibliographystyle{splncs04}
 \bibliography{mybibfile}
\end{document}